\documentclass[submission, PhysProc]{SciPost}
\usepackage[dvipdfmx]{}
\usepackage{here}
\usepackage{amsfonts}
\usepackage{amssymb}
\usepackage{epsfig}
\usepackage{authblk}
\usepackage{braket}
\usepackage{bm}
\usepackage{amsmath}
\usepackage{float}
\usepackage[subrefformat=parens]{subcaption}

\hypersetup{
    colorlinks,
    linkcolor={red!50!black},
    citecolor={blue!50!black},
    urlcolor={blue!80!black}
}

\DeclareSymbolFont{usualmathcal}{OMS}{cmsy}{m}{n}
\DeclareSymbolFontAlphabet{\mathcal}{usualmathcal}

\begin{document}

\begin{center}{\Large \textbf{
Three-body correlations in mesonic-atom-like systems \\
}}\end{center}

\begin{center}
H. Moriya\textsuperscript{1$\star$},
W. Horiuchi\textsuperscript{1} and
J.-M. Richard\textsuperscript{2}
\end{center}

\begin{center}
{\bf 1} Department of Physics, Hokkaido University, Sapporo 060-0810, Japan
\\
{\bf 2} Institut de Physique des 2 Infinis de Lyon, Université de Lyon, CNRS-IN2P3-UCBL,\\
            4, rue Enrico Fermi, Villeurbanne, France
\\

${}^\star$ {\small \sf moriya@nucl.sci.hokudai.ac.jp}
\end{center}

\begin{center}
\today
\end{center}

\definecolor{palegray}{gray}{0.95}
\begin{center}
\colorbox{palegray}{
  \begin{tabular}{rr}
  \begin{minipage}{0.05\textwidth}
    \includegraphics[bb=0 0 308 287,width=14mm]{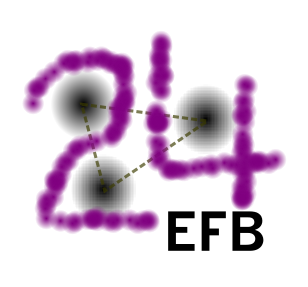}
  \end{minipage}
  &
  \begin{minipage}{0.82\textwidth}
    \begin{center}
    {\it Proceedings for the 24th edition of European Few Body Conference,}\\
    {\it Surrey, UK, 2-4 September 2019} \\
    %\doi{10.21468/SciPostPhysProc.2}\\
    \end{center}
  \end{minipage}
\end{tabular}
}
\end{center}

\section*{Abstract}
{\bf 
Three-body correlations in three-body exotic atoms are studied with simple models
that consist of three bosons interacting through a superposition of
 long- and short-range potentials.
We discuss the correlations among particles by comparing the energy shifts given by precise three-body
 calculations and by the Deser-Trueman formula, 
in which   the  long- and short-range contributions are factorized.
By varying the coupling  of the short-range potential, 
we evaluate the ranges of the strength 
where the two-body correlations dominate
and where the three-body correlations cannot be neglected.}

\vspace{10pt}
\noindent\rule{\textwidth}{1pt}
\tableofcontents\thispagestyle{fancy}
\noindent\rule{\textwidth}{1pt}
\vspace{10pt}

\section{Introduction}
A mesonic atom is a Coulomb bound system consisting of negatively-charged mesons surrounding a nucleus. Studying such systems gives access to the 
properties of the meson-baryon interaction 
at very low energy~\cite{MFLN81,Ikeda11,Ikeda12,MH16,KMOIHOW16,HOHH17,Miyahara18}.
For example, this antikaon-nucleon ($\bar{K}N$) interaction 
is believed to be a strong short-range attraction
as suggested if $\Lambda(1405)$ has a dominant $\bar{K}N$ structure~\cite{KMOIHOW16,AY02}. 
The existence of bound  kaonic nuclei is  still been under discussion
and it is essential to improve our knowledge of  the $\bar{K}N$ interaction~\cite{Ohnishi17}.
The simplest  atom, kaonic hydrogen, consists of 
an antikaon ($K^-$) and a proton ($p$). It was used to 
extract some  information about  
the $\bar K N$ interaction~\cite{BBBBCCCdFF11,BBBBCCCdFF12}.
A study of a kaonic deuterium~\cite{HOHH17,BBBBCCCdFF11,BBBBCCCdFF12} 
gives interesting  constraint on the isospin dependence of the $\bar{K}N$
interaction. 
Encouraged by these results, we investigate whether the physics of exotic atoms can be extended to three-body systems, without assuming that two of them form a nucleus. 
A preliminary  study was made by
one of the present authors (JMR) and C.~Fayard~\cite{RF17}, who
considered a simple system of three identical bosons
interacting via simple long- and short-range potentials. 
By varying  the strength of the short-range term, they studied
the level rearrangement of the spectrum, and
the transition from atomic to nuclear states.
They found that the contributions from
long- and short-range potentials to the energy shifts can be 
factorized within a certain range of the potential strength. Our aim is to extend this study, to consider more
realistic case treated in a  more quantitative manner.

The paper is organized as follows:
In the following sections, we introduce the models, 
the method to solve the three-body problem, and the method of determinant, 
to probe whether the energy shifts are given by a sum of products of 
long- and short-range terms.

%***************************************************************************************************

%*********************************************************************************************************
\section{Models}
\label{sec:model}

In this paper,  two  three-body models are employed.
\subsection{Model I}

The simplest model consists of  three identical bosons. 
All interactions between two particles 
have long-range and short-range attraction parts. 
The Hamiltonian of this system is
\begin{equation}
	H_{\mathrm{I}} = \sum_{i=1}^{3} T_{i} - T_{cm} + \sum_{i>j=1}^{3} V_{ij}^{LR} +  \lambda  \sum_{i>j=1}^{3} V^{SR}_{ij},
\end{equation}
where $T_{i}\;(i=1,2,3)$ is the kinetic energy of the $i$th particle 
and $T_{cm}$ is the kinetic energy of the center of mass, which is subtracted. 
All the physical constants including masses are set to 1. 
The long-range ($LR$) and short-range ($SR$) two-body potentials 
have only a central term.
The strength of the short-range potential is varied
through  the parameter~$\lambda$. 
We assume a regularized Coulomb for the long-range part and a Gaussian shape for the 
short-range potential. The explicit forms are
\begin{align}
	&V^{LR}_{ij} = -\frac{\mathrm{erf}(\mu_{LR}\,r_{ij})}{r_{ij}} \label{eq:lrm1}, \\
	&V^{SR}_{ij} = -C_{SR}\,\mu ^{3}_{SR} \exp \left(- \mu_{SR}^{2}\, r_{ij}^{2}\right) ,\label{eq:srm1}
\end{align}
where $r_{ij}$ denotes the distance between the $i$th and $j$th particles. 
The strength parameter $C_{SR}$ is tuned so 
that the the short-range potential alone supports  a two-body bound state for $\lambda>1$, i.e., $\lambda=1$ is the coupling threshold for binding. 
The range parameters of both the short-range potential and the regularizing term of the long-range potential
are set to $\mu_{LR}=\mu_{SR}=30$, which is large compared to the inverse Bohr radius, 
so that the role of the long- and short-range interactions
are well separated. Since all the long-range interactions are attractive,
this model cannot be realized by Coulombic systems,
it corresponds to a gravitational interaction.

\subsection{Model II}

\begin{figure}
	\begin{center}
		\includegraphics[width=.8\linewidth,bb=0.000000 0.000000 1119.057507 635.032634]{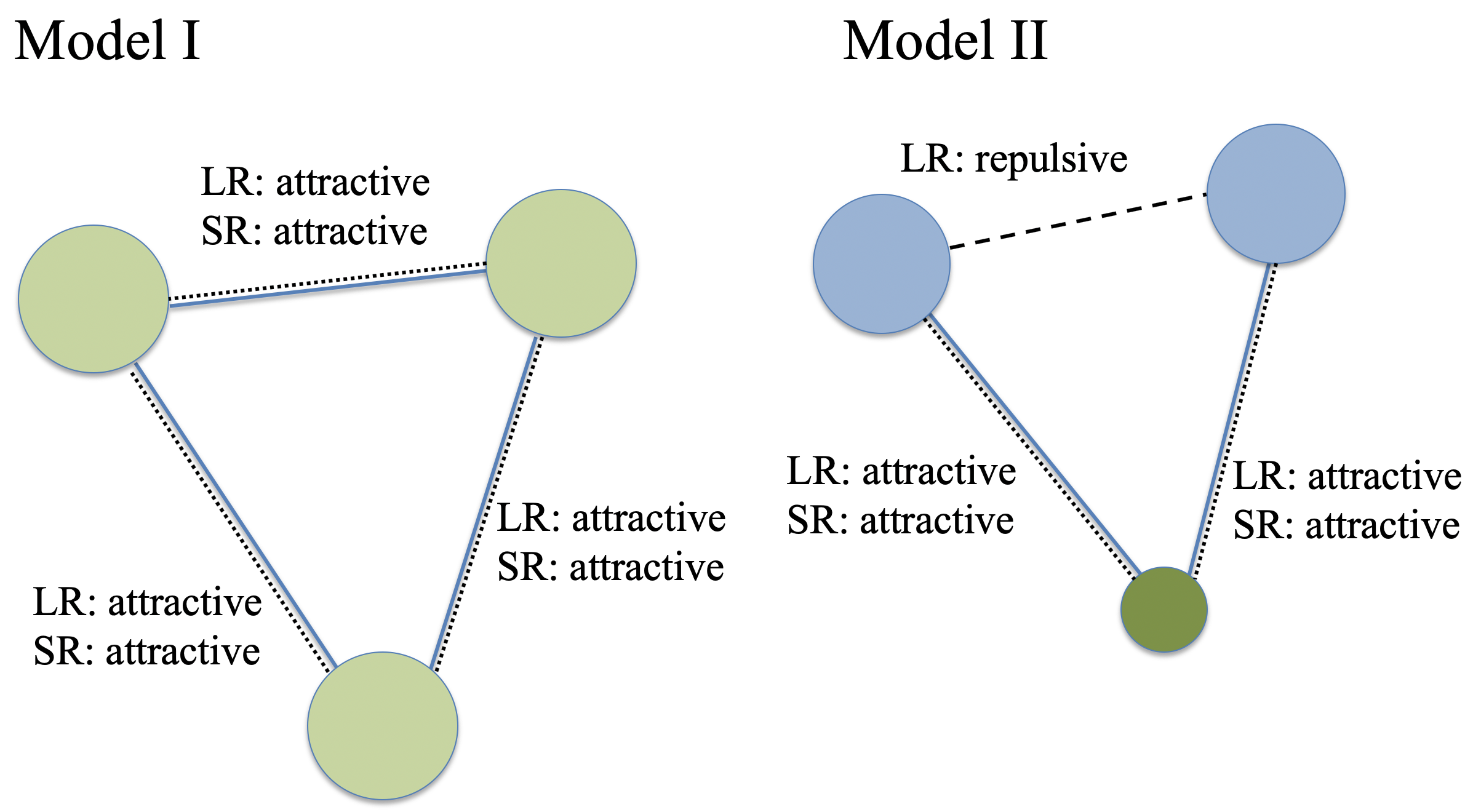}
		\caption{Schematic pictures of Model I and II employed in this paper. 
		Solid lines represent the short-range attractive potentials, and dotted and dashed lines represent the long-range attractive and repulsive potentials, respectively.}
	\label{model_schematic}
	\end{center}
\end{figure}

Model II describes  a  case that is more realistic, or at least closer to the $ppK^{-}$ system. 
The first and second particles 
are identical bosons with a mass $m_{1}=m_{2}=1$ and a positive charge $q_1=q_2=+1$, 
while the  third particle, also spinless, has a mass $m_{3}=1/2$ and a charge $q_3=-1$.
The short-range potential is restricted to the interaction with the third particle,
with $C_{SR}$ appropriately rescaled so that $\lambda=1$ is the coupling threshold for a two-body system of masses $\{m_1,m_3\}$. 
The Hamiltonian of Model II is
\begin{equation}
	H_{\mathrm{II}} = \sum_{i=1}^{3} T_{i} - T_{cm} +\sum_{i>j=1}^{3} V_{ij}^{LR}  +  \lambda\sum_{i=1}^{2} V^{SR}_{i3}
\end{equation}
with
\begin{align}
	&V^{LR}_{ij} =q_i\,q_j\, \frac{\mathrm{erf}(\mu_{LR}r_{ij})}{r_{ij}},  \\
	&V^{SR}_{ij} = - C_{SR}\, \mu_{SR} ^{3} \exp \left(- \mu_{SR}^2\, r_{ij}^2\right). 
\end{align}
Model I and Model II  
are schematically summarized in Fig.~\ref{model_schematic}.

%*******************************************************************************************************
\section{Correlated Gaussian expansion}
\label{sec:Tc}
The three-body calculations are carried out by a well-known variational method, which is now briefly summarized. 
Let $\bm{x}$ denote the set relative coordinates,
\begin{equation}
	\bm{x} = \begin{pmatrix}
				\bm{x}_{1} \\
				\bm{x}_{2}
				\end{pmatrix}.
\end{equation}
Here we choose the Jacobi coordinates:
\begin{align}
		&  \bm{x}_{1} = \bm{r}_{1}-\bm{r}_{2}~,   \\
		&  \bm{x}_{2} = \frac{m_{1}\bm{r}_{1} + m_{2}\bm{r}_{2}}{m_{1}+m_{2}}-\bm{r}_{3},
\end{align}
where $\bm{r}_{i}$ $(i=1,2,3)$ is the single-particle coordinate of the $i$th particle.
The three-body wave function $\ket{\Psi^{(3)}}$ 
is expanded on a basis of  correlated Gaussians (CG) \cite{VS95},
\begin{equation}
\ket{\Psi^{(3)}} = \sum_{k} c_{k} \mathcal{S} \exp \left( -\frac{1}{2} \tilde{\bm{x}} A_{k} \bm{x} \right),
\end{equation}
where $\mathcal{S}$ is symmetrizer acting on the three particles (Model I) or on the $\{1,2\}$ subset (Model II), and 
$A_{k}$ is the positive-definite 2$\times$2 symmetric real matrix 
which characterizes the $k$th CG.
The energy and the expansion coefficients $\{c_{k}\}$ are determined by
solving a generalized eigenvalue problem.
To optimize the non-linear variational parameters entering  the $A_{k}$,
we employ the stochastic variational method~\cite{VS95,SVMbook}.
Since we have  to treat simultaneously two different scales, atomic and nuclear, 
we adopt  the following strategy in the search for the variational parameters.
Suppose that we have already a basis of  $K$ CG:
A number of candidates for the  additional $A_{K+1}$ matrices are generated randomly 
with their elements either at  the nuclear or  atomic scale.
For small $K$, we select the matrix  providing the minimum energy.
Once the energy is converged up to  a certain number of digits,
the additional CG are generated only with elements at the nuclear scale.
This procedure is efficient, particularly with large $\lambda$,
where the wave function changes drastically at short distances.
In our calculations, we have increased the size of the basis until the energy is converged
within $10^{-4}$.

\section{Factorization of the long- and short-range contributions}
%
%*******************************************************************************************************
\subsection{Deser-Trueman formula}
\label{sec:DT}
\begin{figure}[H]
\centering
\includegraphics[width=8cm,bb=0.000000 0.000000 235.510938 226.160938]{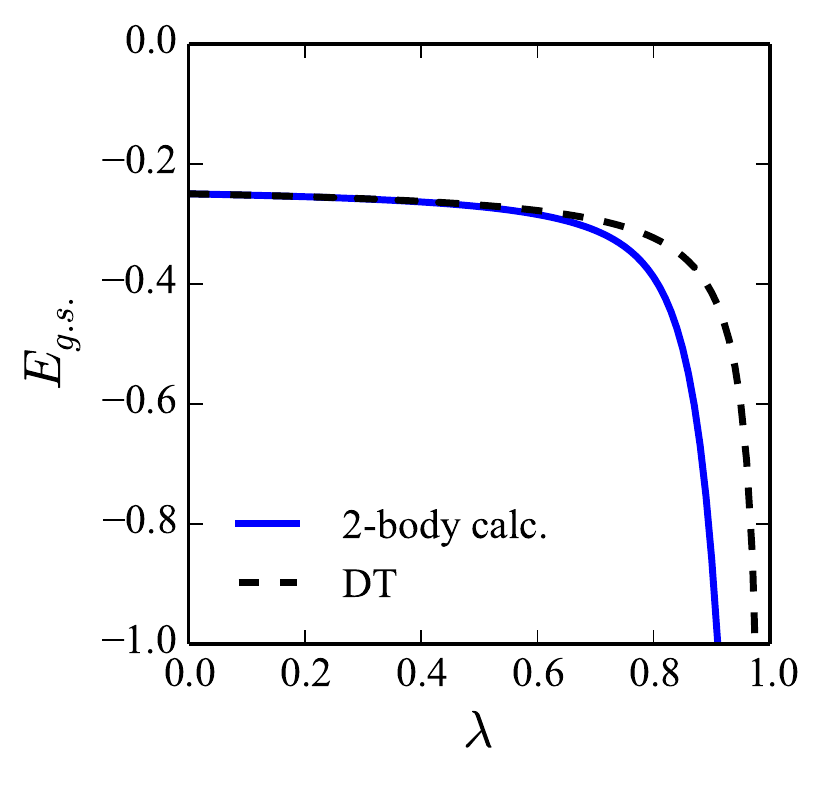} 
\caption{Comparison of the two-body ground-state energy, $E_{g.s.}$, obtained with the full two-body calculation
and the DT formula.}
\label{2b}
\end{figure}

The energy shift of two-body exotic atoms is often  estimated 
with the Deser-Trueman (DT) formula \cite{DGBT54,T61}. 
\begin{equation}
	\delta E^{(2)} = \frac{2 \pi}{\mu} |\Psi^{(2)}_{0}(0)|^{2} a,
\end{equation}
where $\mu$ is the reduced mass, 
$\Psi^{(2)}_{0}(0)$ is the relative wave function at the origin obtained
with the long-range potential alone, and
$a$ is the scattering length calculated with by the short-range potential alone.
Note the remarkable factorization of the long-range and short-range contributions in the DT formula.
Fig.~\ref{2b} shows a comparison of the ground-state energy
of  two identical bosons  interacting with 
Eqs.~(\ref{eq:lrm1}) and (\ref{eq:srm1}), calculated either exactly or 
by the DT formula, 
with the strength $\lambda$ of the short-range potential  varied continuously.  
At small $\lambda$, the DT formula reproduces well the ground-state energy
obtained by direct  two-body calculations.
However, the DT formula deviates
from the two-body calculation as $\lambda$ increases.
This shows that the energy shift involves higher-order corrections, beyond the simple scattering length in the DT formula.
A parallel question is whether or not the energy shift can 
still  be factorized into
the long- and short-range contributions in large $\lambda$ region. 
For more quantitative discussion, we introduce in the next subsection the determinant
method.

\subsection{Determinant Method}
\label{sec:dm}
To evaluate quantitatively the validity of the factorization of the energy shift, 
we use the following method. 
Let $M$ be the matrix of the energy-shifts  for a series of discretized strengths $\lambda_1, \lambda_2,\dots$ 
and several long-range potentials, as spelled out in  Table~\ref{matM.tab}.
\begin{table}[H]
\caption{Matrix $M$ constructed from the energy shifts
obtained with different long- and short-range potentials.}
\label{matM.tab}
\begin{center}
	\begin{tabular}{c||c|c|c|c}
			& $\lambda_{1}$ & $\lambda_{2}$ & $\lambda_{3}$ & $\cdots$  \\ \hline \hline
		$LR_{\mathrm{I}}$ & $\delta E (LR_{\mathrm{I}},\lambda_{1})$ & $\delta E (LR_{\mathrm{I}},\lambda_{2})$ & $\delta E (LR_{\mathrm{I}},\lambda_{3})$ & $\cdots$  \\ \hline
		$LR_{\mathrm{I\hspace{-.1em}I }}$ & $\delta E (LR_{\mathrm{I\hspace{-.1em}I} },\lambda_{1})$ & $\delta E (LR_{\mathrm{I\hspace{-.1em}I} },\lambda_{2})$ & $\delta E (LR_{\mathrm{I\hspace{-.1em}I} },\lambda_{3})$ & $\cdots$  \\ \hline
		$LR_{\mathrm{I\hspace{-.1em}I\hspace{-.1em}I} }$ & $\delta E (LR_{\mathrm{I\hspace{-.1em}I\hspace{-.1em}I} },\lambda_{1})$ & $\delta E (LR_{\mathrm{I\hspace{-.1em}I\hspace{-.1em}I} },\lambda_{2})$ & $\delta E (LR_{\mathrm{I\hspace{-.1em}I\hspace{-.1em}I} },\lambda_{3})$  & $\cdots$  \\ \hline
		$\vdots$ & $\vdots$ & $\vdots$ & $\vdots$ &$\ddots$ \\
	\end{tabular}
\end{center}
\end{table} 
If the level shift can be factorized as the product of  a contribution
from the long-range potential and another from the  short-range part potential,
as in the DT formula, 
the determinant of any $2\times 2$ submatrix $S_{2}$ taken from $M$
must be zero. 
For example, for $\lambda_{i}$ when $\delta E(LR, \lambda_{i})$ 
is the product of separated contributions 
from the long-range and short-range interactions, 
that is $\delta E(LR, \lambda_{i}) = A_{LR}B_{SR}(\lambda_{i})$. 
The submatrix $S_{2}$ is defined by
\begin{equation}
	S_{2} = \begin{pmatrix}
				\delta E (LR_{\mathrm{I}},\lambda_{1})   &   \delta E (LR_{\mathrm{I}},\lambda_{2})       \\
				\delta E (LR_{\mathrm{I\hspace{-.1em}I} },\lambda_{1})    &    \delta E (LR_{\mathrm{I\hspace{-.1em}I} },\lambda_{2})
			\end{pmatrix} 
			=\begin{pmatrix}
				A_{LR_{\mathrm{I}}}B_{SR}(\lambda_{i})   &   A_{LR_{\mathrm{I}}}B_{SR}(\lambda_{i+1})     \\  
				A_{LR_{\mathrm{I\hspace{-.1em}I}}}B_{SR}(\lambda_{i})   &   A_{LR_{\mathrm{I\hspace{-.1em}I}}}B_{SR}(\lambda_{i+1})     \\  
			\end{pmatrix}.
\end{equation}
Considering the determinant of $S_{2}$, it can be easily proven that the $\mathrm{det}S_{2}$ is zero analytically as
\begin{equation}
	\mathrm{det}S_{2} = A_{LR_{\mathrm{I}}}B_{SR}(\lambda_{i})A_{LR_{\mathrm{I\hspace{-.1em}I}}}B_{SR}(\lambda_{i+1}) - A_{LR_{\mathrm{I}}}B_{SR}(\lambda_{i+1})A_{LR_{\mathrm{I\hspace{-.1em}I}}}B_{SR}(\lambda_{i}) = 0
\end{equation}
On contrary, when the energy shift is not separable, 
then $\mathrm{det}S_{2}$ is not necessarily zero.
Practically, to get the variation of the potentials,
we take the two long-range potentials with $\mu_{LR}=10$ and $\mu_{LR}=30$ 
and different $\lambda$s at intervals of 0.01 ($\lambda_{i+1}-\lambda_{i}=0.01$).

\subsection{Factorization of long- and short-range contributions in 
two-body system}
\begin{figure}[H]
\centering
%%PDFVersion: 1.4
\includegraphics[width=8cm, bb=0.000000 0.000000 239.385938 220.715625]{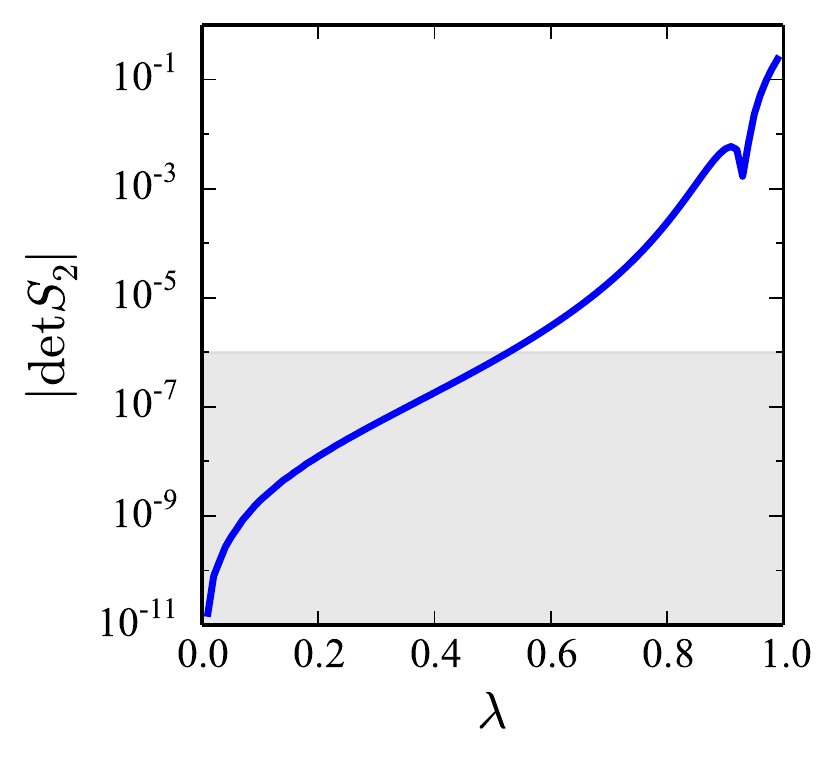} 
\caption{The determinant of the $S_{2}$ of the two-body system.
The shaded region shows the region that in which $\det S_2$ can be considered to be zero. }
\label{det2b}
\end{figure}

Let us show how the determinant method works for the two-body system.
Figure~\ref{det2b} plots $|\det S_2|$ as a function of
the strength of the short-range potential $\lambda$. 
To appreciate what $\det S_2\simeq 0$ means, we take into account the order of magnitude of the elements of $S_2$ and  the accuracy of the calculation.  In Fig.~2b, this  corresponds to $|\det S_2|\lesssim 10^{-6}$. 
The shaded area in Fig.~\ref{det2b} indicates 
the possible regions where the numerical error dominates
Here the range of the strength $\lambda$ where the factorization holds is seen to be about 
 $\lambda\lesssim \lambda_{c}=0.6$. Interestingly this is the range of $\lambda$ for which the DT formula works very well. 

%
%**************************************************************************************************
\section{Discussions: Three-body correlations}
To discuss the results of three-body models, the two-body DT formula
is extended to the three-body case as \cite{RF17}
\begin{equation}
        \sum_{i>j=1}^{3}\frac{2\pi}{\mu_{ij}} |\Psi_{0,ij}^{(3)}(0)|^{2}a_{ij}, 
\end{equation}
where $\mu_{ij}$, $a_{ij}$ are respectively the reduced mass and the
scattering length obtained only with the short-range potential
of the $i$th and $j$th particles,
and $|\Psi^{(3)}_{0,ij}(0)|^{2}$ is defined by 
\begin{equation}
	|\Psi^{(3)}_{0,ij}(0)|^{2}=\frac{\braket{\Psi^{(3)}_{0}|\delta(\bm{r}_{i}-\bm{r}_{j})|\Psi^{(3)}_{0}}}{\braket{\Psi^{(3)}_{0}|\Psi^{(3)}_{0}}}.
\label{extDT.eq}
\end{equation}
$\Psi^{(3)}_{0}$ is the wave function obtained 
only with the long-range potential.
Note that the extended DT formula keep the form 
of a sum of  products of contributions 
from the long- and short-range potentials.

\begin{figure}[!t]
	\begin{tabular}{c}
	\begin{minipage}{0.5\hsize}
	%\centering
	\includegraphics[width=6.5cm,bb=0.000000 0.000000 235.510938 404.720937]{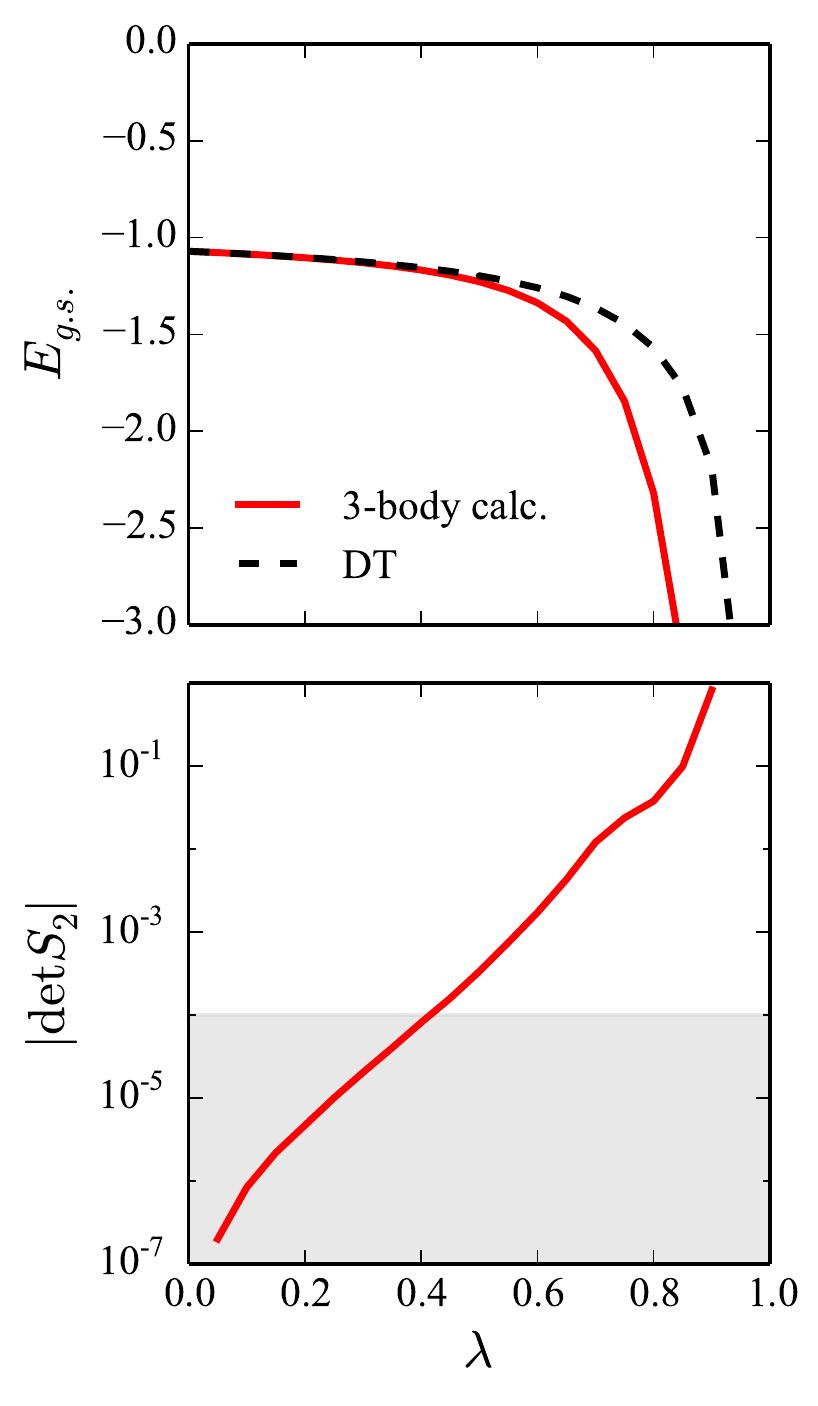} 
	\subcaption{Model I}
	\label{3bI}
	\end{minipage}
	%***************************
	\begin{minipage}{0.5\hsize}
	%\centering
	\includegraphics[width=6.5cm,bb=0.000000 0.000000 235.510938 404.720937]{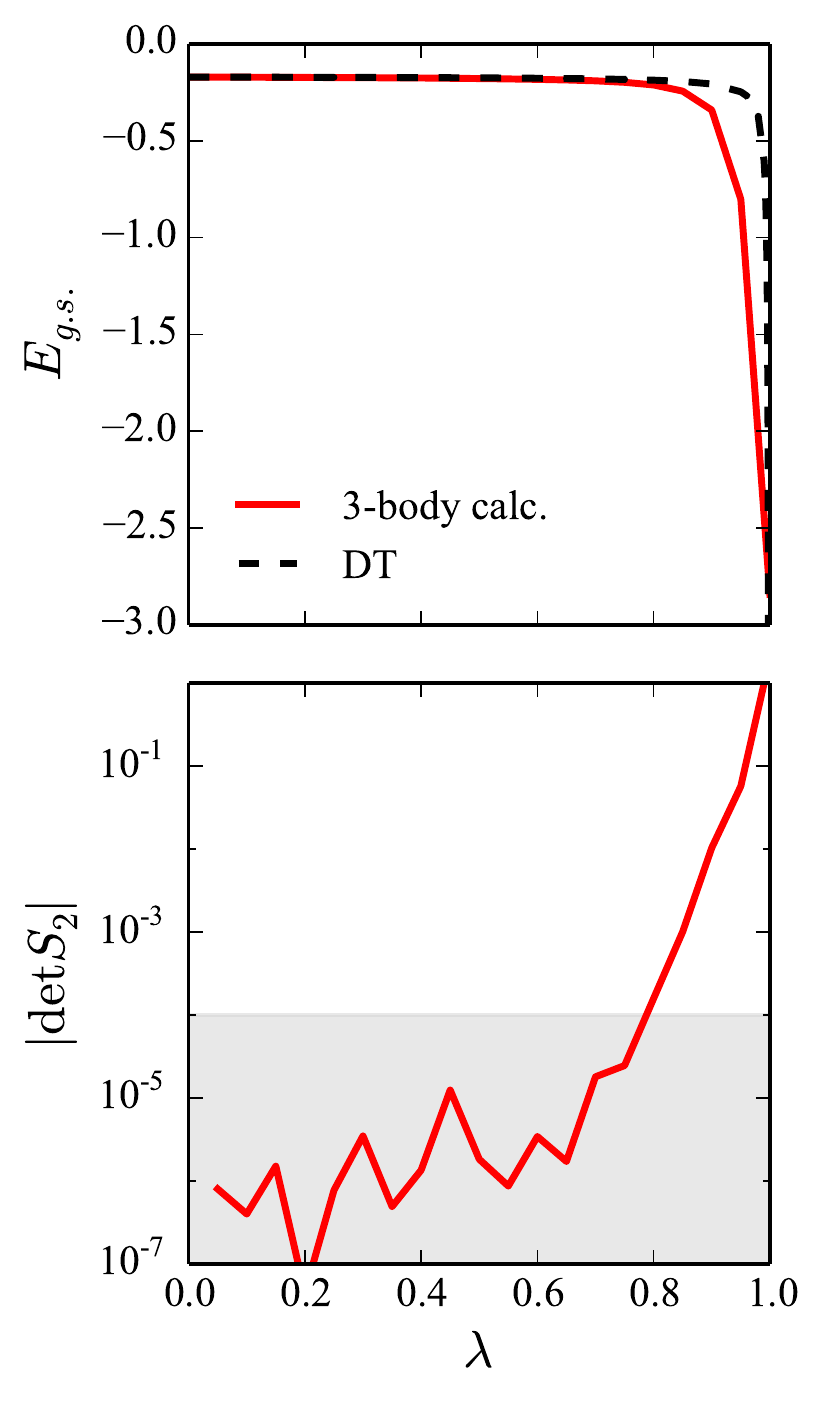} 
	\subcaption{Model II}
	\label{3bII}
	\end{minipage}
	\end{tabular}
	\caption{(Upper) Ground-state energy $E_{g.s.}$ obtained 
by the full three-body calculation and the DT formula.
(Lower) $|\det S_2|$ values calculated with the three-body calculations.}
\end{figure}

The upper panel of Fig.~\ref{3bI} shows 
comparison between the ground-state energy obtained by the 
full three-body calculation and the extended DT formula 
of Eq.~(\ref{extDT.eq}) for Model I.
At small values of $\lambda$, the energy shift is small and shows a flat behavior. The extended DT formula reproduces well the energy shift
of the three-body calculation in this flat region. From the lower left panel, one can see that the factorization is also  valid in that region. 
Then the energy shift drops rapidly at some $\lambda$, and the factorization breaks down simultaneously (we again estimated the area for which a vanishing of the determinant makes sense, given the order of magnitude of the matrix elements and the accuracy of the calculation).  A departure for the DT approximation is observed at about $\lambda_c\simeq 0.4$, while in the two-body case, a similar departure occurred only at $\lambda_c\simeq 0.6$. This is because in the latter case, a purely nuclear state requires $\lambda=1$, for which $a\to \infty$, while in the former case, a Borromean three-body bound state occurs for $\lambda\simeq0.8$. Hence the atomic spectrum is ``pulled down'' earlier. 

Figure~\ref{3bII} displays the same plots  for Model II.
The energy shift and the determinant
exhibit  the same qualitative behavior but
the critical strength becomes much larger, $\lambda_c\simeq  0.8$. 
This is because  of the repulsive long-range potential 
between two identical bosons, which suppresses 
the three-body correlations. The level shift of such a three-body system
is determined only by the pairwise correlations and is of factorizable form. 

%*************************************************************************************************
\section{Summary}
Accurate three-body calculations have been performed to evaluate three-body correlations in exotic-atom-like three-body systems.  The interaction, which is pairwise, consists of a Coulomb-type of long-range interaction and a short-range potential whose strength is varied. Two models have been considered. 
Model I consists of three identical bosons.
Model II includes two identical bosons of mass $m_{1,2}=1$ and a third particle of mass $m_3=1/2$, and opposite charge. 
The factorization property of the long- and short-range contributions
to the energy shift have been examined quantitatively by the determinant method.

We find that, when the strength of the nuclear  interaction is increased,   the factorization and the dominance of two-body correlations break down earlier
when the same long- and short-range potentials are applied 
to all pairs (Model I), whereas the three-body correlations
are much smaller with Model II in which only two pairs interact. 
This is intimately related to the early or delayed occurrence of a Borromean three-body bound state in the nuclear potential. 

For further extension of this study, 
the analysis of the excited states is underway 
for a general understanding of the many-body correlations
and of the level rearrangement.  
In particular, we shall extend the method of the determinant 
to larger submatrices to probe whether the energy shift 
is a sum of products of long- and short-range terms, rather than a mere product. 
We also aim at investigating such exotic-atom-like
systems with a complex potential to take the meson-baryon 
absorption effect into account. This is, indeed, an important aspect of 
the $\bar{K}N$ interaction~\cite{AY02,MH16}. 

\section*{Acknowledgements}
We acknowledge the collaborative research program 2019,
information initiative center, Hokkaido University.

\paragraph{Funding information}
This work was in part supported by JSPS
KAKENHI Grants No. 18K03635, No. 18H04569, and No. 19H05140.

\bibliographystyle{SciPost_bibstyle} % Include this style file here only if you are not using our template

\end{document}